\documentclass[prl, twocolumn, floatfix]{revtex4}
\setcitestyle{super}

\usepackage{graphicx}
\usepackage{hyperref}
\usepackage{amsmath}
\usepackage{color}

\begin{document}

\title{Analysing spatiotemporal instabilities in magneto-optical traps with the tools of turbulence theory}

\author{Adam Griffin$^{1}$, M. Gaudesius$^{1,2}$, R.\ Kaiser$^{1}$, Sergey Nazarenko$^{1}$ and G.\ Labeyrie$^{1}$\footnote{To whom correspondence should be addressed.}}
\affiliation{$^{1}$Universit\'{e} C\^{o}te d'Azur, CNRS, Institut de Physique de Nice, 06560 Valbonne, France}
\affiliation{$^{2}$Department of Physics and Astronomy, University of Oklahoma, Norman, Oklahoma 73019, USA}

\date{\today}
\begin{abstract} A large cloud of $^{87}$Rb atoms confined in a magneto-optical trap exhibits, in a certain regime of parameters, spatio-temporal instabilities with a dynamics resembling that of a turbulent fluid. We apply the methods of turbulence theory based on structure function analysis to extract scaling exponents which are compared to known turbulent regimes. This analysis also allows us to make a clear distinction between different instability regimes.
\end{abstract} 
\maketitle

\section{Introduction}

Fluid properties of nonlinear light systems and Bose-Einstein condensates (BEC) have been noticed and  scrutinized over the past two or three decades. These studies have been particularly well advanced for systems described by the Gross-Pitaevskii (nonlinear Sch\"odinger) equation for which the Madelung transformation (the amplitude-phase decomposition of the complex field) allows to bring this equation to the fluid mass and momentum conservation form. Among these studies, works devoted to turbulence in optical~\cite{Dyachenko1992} and BEC systems~\cite{Bagnato2009, Anderson2013} are most interesting and important. 

However, there are also optical systems which seem to exhibit turbulent behaviors, but whose dynamical equations are not fully known or too complicated to be analyzed analytically or even numerically. Yet, as we will show in this work in the case of a magneto-optical trap (MOT), a lot can be said about such systems using the standard characterizations from the turbulence theory.

In this work, we will analyze the data obtained from an unstable MOT experiment. Unstable MOTs have been studied in various groups~\cite{Wilkowski2000, diStefano2003, Labeyrie2006, Romain2016, Gaudesius2020, Gaudesius2021, Giampaoli2021}, with different models ranging from atomic physics~\cite{Pohl2006}, non linear dynamics~\cite{Stefano2004} to plasma physics~\cite{Mendonca2008} and astrophysics~\cite{Mendonca2012}. In all instances, the experimental data were obtained from temporal and/or spatial analysis of fluorescence or absorption images of unstable clouds using tools such as statistical analysis~\cite{Gaudesius2021}, Principal Component Analysis~\cite{Romain2016} or autocorrelation functions~\cite{Giampaoli2021}. In the present work we employ, for the first time to our knowledge in this context, a method based on the analysis of structure functions (SF) commonly used in the study of turbulent systems. This method is in principal far more powerful, as it gives access to scaling exponents for all involved spatial scales. For instance, one is in principle able to detect small-scale turbulent fluctuations on a large mean-field background.

Our approach is motivated by the fact that in some range of parameters our MOT exhibits spatio-temporal fluctuations that are visually reminiscent of turbulence, although we stress that we do not possess a theoretical model to confirm this observation. However, we note that in the work of Ref.~\cite{Giampaoli2021} using a MOT very similar to ours, a simplified model of diffusive light transport coupled to atomic density via radiation pressure was employed to interpret the experimental observations as ''photon bubble turbulence''~\cite{Mendonca2012}. Even though the analysis presented in this Letter is purely based on data treatment, we will see that it allows to identify different instability regimes for the MOT, in qualitative agreement with our previous findings~\cite{Gaudesius2021}.

\begin{figure}
\begin{center}
\includegraphics[width=1.0\columnwidth]{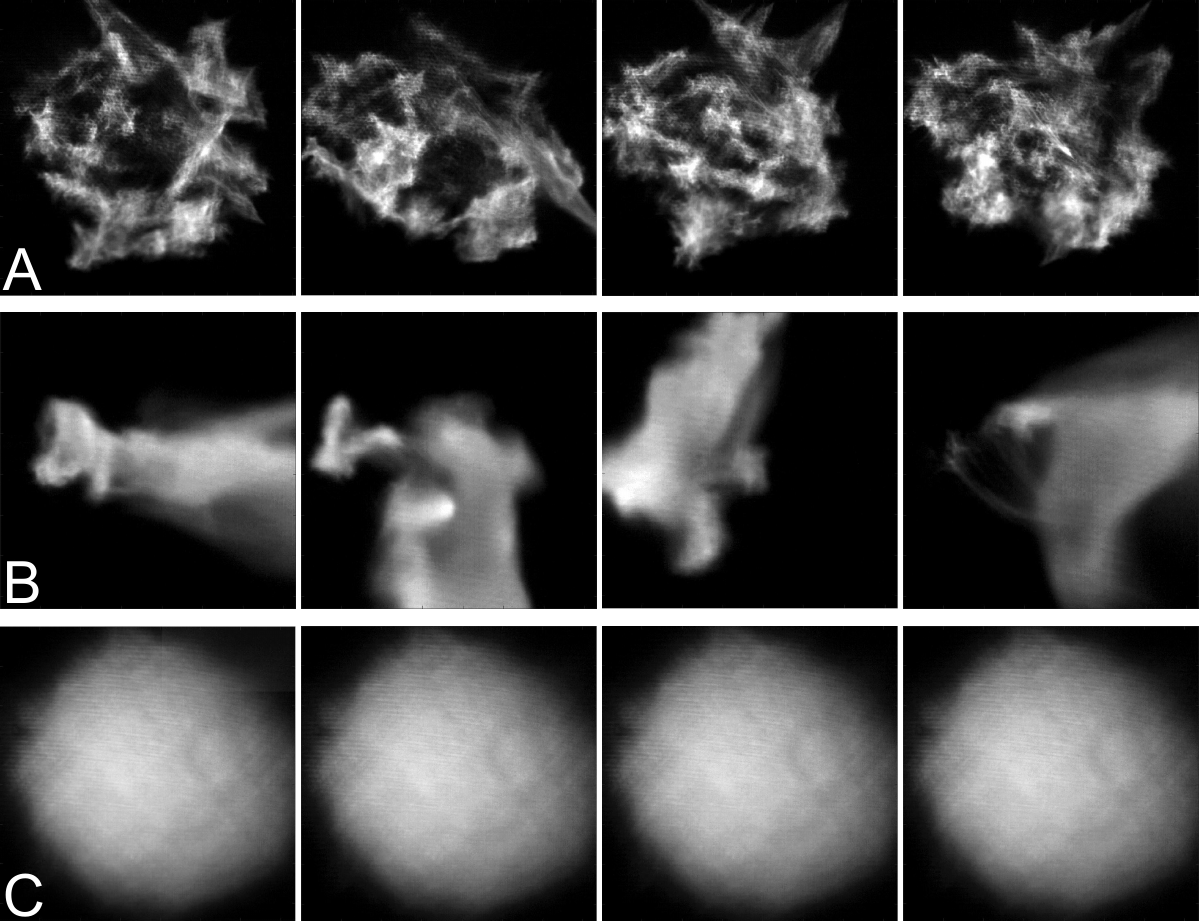}
\caption{Illustration of different MOT regimes. We show examples of single-shot MOT fluorescence images recorded at random time. Each row correspond to different MOT parameters. A (unstable MOT): $\nabla$B = 1.2 G/cm, $\delta$ = -$\Gamma$. B (unstable MOT): $\nabla$B = 12 G/cm, $\delta$ = -$\Gamma$. C (stable MOT): $\nabla$B = 2.4 G/cm, $\delta$ = -4$\Gamma$. The frame size is that used for the computation of the structure function, and is different for each row (see text).}
\label{fig1}
\end{center}
\end{figure}

\section{Structure function} 

In this work, we will employ the structure function which is the most commonly used object in turbulence analysis~\cite{Frisch1995}. Considering a given field $\rho(r)$, for instance a velocity field, the structure function of order $p$ is defined by:
\begin{equation}
SF_p(\ell) = \langle | \rho(\boldsymbol r_1) - \rho(\boldsymbol r_2) |^p \rangle
\end{equation}
where $\ell$ is the distance between points $\boldsymbol r_1$ and $\boldsymbol r_2$, and the brackets denote averaging over space. 
In turbulence, the structure function exhibits a scaling behavior, i.e. it  behaves as a power law $SF_p(\ell) \propto \ell^{\xi_p}$ in a wide range of $\ell$. This scaling range is often referred to as the inertial range. The quantities ${\xi_p}$ are called the structure function exponents. They contain significant informations about the turbulence statistics, typical processes (e.g. turbulent cascades) and presence of coherent and/or singular structures causing intermittency of the turbulent signal (e.g. shocks). We provide in the following a few examples in the case of known regimes:

\textit{Differentiable fields} (often called smooth ramps): $\xi_p =p$. This result follows from a Taylor expansion of the density profile at small $\ell$:
$\left|\rho(r+\ell)-\rho(r)\right|^p = \frac{d\rho}{dr}^p*\ell^p \propto \ell^p$.

\textit{Shocks}: $\xi_p$ = 1. This result relies on the fact that for sufficiently small $\ell$, the main contribution to $SF_p(\ell)$ comes from pairs of points located on each side of the sharp interface of the shock, whose proportion grows like $\ell$ hence $SF_p(\ell) \propto \ell$.

\textit{Burgulence} (random field governed by the Burgers equation). Here smooth ramps coexist with shocks. Hence, a bifractal behaviour is observed~\cite{Frisch1995}: ramp scaling is seen for $0<p<1$ and shocks for $1 < p < \infty$.

\textit{Passive scalar advected by Kolmogorov turbulence}. This case approximately describes the temperature field in atmospheric turbulence. The mean-field Kolmogorov-Obukhov-Corssin (KOC) theory~\cite{Kolmogorov1941,Obukhov1949,Corrsin1951} predicts ${\xi_p} =p/3$ for the scaling range $\ell_d < \ell < \ell_E$, where $\ell_d$ and $\ell_E$ are the dissipative and the energy-containing (integral) scales respectively. The scaling is derived under the assumption that the energy dissipation rate is the only quantity defining the statistical properties in this range.

\textit{Intermittent turbulence}. This case usually corresponds to real turbulent systems: Kolmogorov-type scaling at low values of $p$ is replaced by a more shallow slope for higher $p$'s due to presence of coherent quasi-singular structures/events~\cite{Leveque1999,Iyer2018}.

\begin{figure}
\begin{center}
\includegraphics[width=1.0\columnwidth]{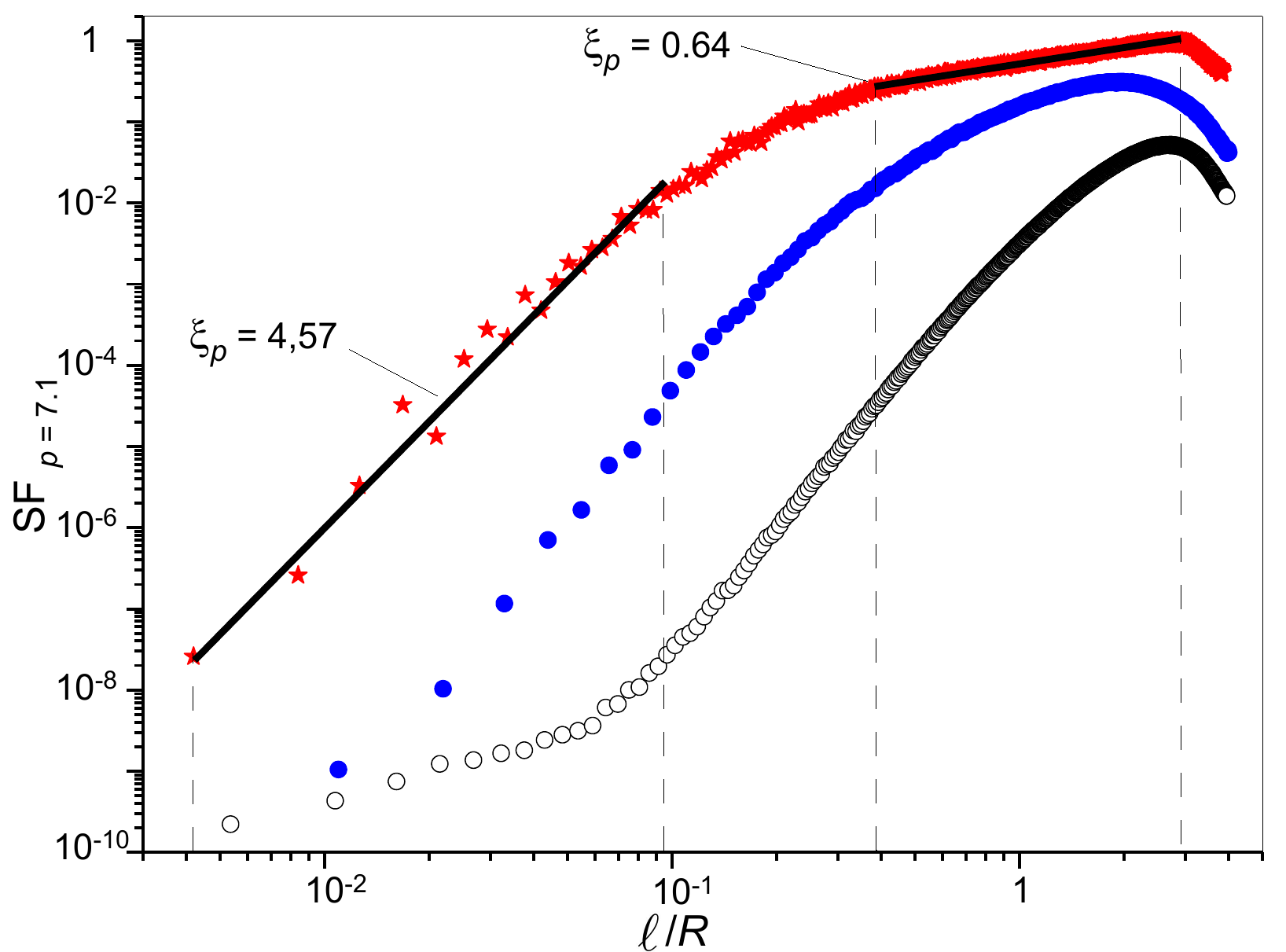}
\caption{Examples of structure functions. We plot here three examples of structure functions, obtained from the data of Fig.1 with $p = 7.1$ (note the log-log scale). The horizontal scale is for each case normalized to $R$. Stars: $\nabla$B = 1.2 G/cm, $\delta$ = -$\Gamma$. Dots: $\nabla$B = 12 G/cm, $\delta$ = -$\Gamma$. Circles: $\nabla$B = 2.4 G/cm, $\delta$ = -4$\Gamma$. In the first case, two clear scaling ranges can be observed (dashed lines), with different scaling exponents.}
\label{fig2}
\end{center}
\end{figure}

\section{Experimental Setup and Data treatment} 

We use a MOT setup able to trap and cool a large number $N$ of $^{87}$Rb atoms ($N$ up to $1.5 \times 10^{11}$, see~\cite{Camara2014}). The resulting cloud of atoms is centimeter-sized, with a temperature around $200~\mu$K (in the stable regime). For such large $N$ values, the MOT is known to exhibit spatio-temporal instabilities when the trapping laser frequency is brought sufficiently close to the atomic transition frequency~\cite{Labeyrie2006, Pohl2006, Gaudesius2020, Rodrigues2022}. This is due to competing collective forces arising from multiple scattering of light inside the atomic cloud.

In a previous study~\cite{Gaudesius2021}, we have shown that different unstable regimes could be identified, depending on experimental parameters such as the laser detuning $\delta~=~\omega_{L}~-~\omega_{at}$ (where $\omega_{L}$ and $\omega_{at}$ are the laser's and atomic frequencies respectively) and the magnetic field gradient $\nabla B$. This is illustrated in Fig.1, where we show single-shot fluorescence images recorded by a CCD camera at random times (see below), for three sets of MOT parameters: (A) $\nabla B = 1.2$ G/cm and $\delta = -\Gamma$; (B) $\nabla B = 12$ G/cm and $\delta = -\Gamma$; and (C) $\nabla B = 2.4$ G/cm and $\delta = -4\Gamma$, where $\Gamma$ is the width of the atomic transition. The latest corresponds to a stable cloud, used as a reference. We provide in the Supplementary Material time-resolved videos corresponding to these three situations. In the present work, we will compare several unstable clouds corresponding to different values of $\nabla B$ (1.2, 1.7, 2.4, 4.8, 7.2, 9.6 and 12 G/cm) and a fixed detuning $\delta = -\Gamma$.

\begin{figure}
\begin{center}
\includegraphics[width=1.0\columnwidth]{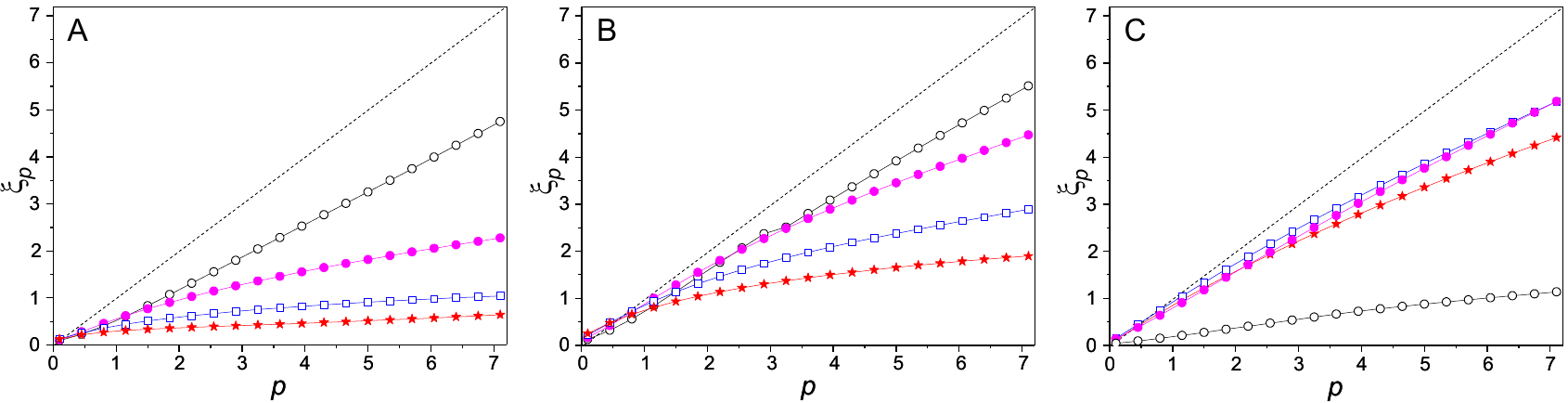}
\caption{Behaviors of structure function exponents. We report on this graph the evolution of $\xi_p$ obtained for different scaling ranges: $\xi_p^1$ for $2 \leq \ell \leq 10$ (black circles), $\xi_p^2$ for $0.1R \leq \ell \leq 0.3R$ (magenta dots), $\xi_p^3$ for $0.3R \leq \ell \leq 0.6R$ (blue open squares), and $\xi_p^4$ for $0.6R \leq \ell \leq 1.2R$ (red stars). The dotted line corresponds to the case of a smooth field ($\xi_p = p$). The different panels correspond to the data in Fig.1.}
\label{fig3}
\end{center}
\end{figure}

To compute the SF, we use a set of 100 fluorescence images of the cloud, collected by the CCD camera at random times during the dynamics. The exposure time for each image is 1 ms. We first compute the average image by summing over the full data set. This allows us to determine the center of mass (\textit{com}) of this average image, and its \textit{rms} radius $R$. We then define a square window of width $L$, centered on the \textit{com} of the average image. In the following, we use $L = 4 \times R$. For each image of the data set, we compute the SF by randomly choosing many pairs of points within the window. The SF corresponding to all images are then averaged. We thus perform an averaging over space and time, assuming ergodicity and spatial isotropy. Since the MOT imaging takes place in the plane transverse to the axis of the coils producing the magnetic field gradient, we expect the last hypothesis to be reasonable.

Various finite-size and smoothing effects can affect the SF and reduce the effective size of scaling ranges. The minimal size accessible in our imaging is the pixel size ($\approx 30~\mu$m). The upper limit of spatial scales is set by the cloud's radius $R$, which depends on $\nabla B$. The finite exposure time of the CCD results in a smoothing of the small-scale structures in the images. For the mean atomic velocity in a stable cloud (0.1 m/s), this smoothing effect is expected to occur for spatial scales below $100~\mu$m (approximately three pixels on the CCD). Note that for highly unstable clouds (i.e. for large values of $\nabla$B), atomic velocities may be quite larger resulting in a degraded resolution. In the following, we assume that the 3D light intensity distribution reflects the atomic spatial density distribution. Note that this is only an approximation, since radiation trapping effect are known to deform the spatial distribution of scattered light~\cite{Camara2014}. Furthermore, the recorded 2D images correspond to the projection of the 3D fluorescence light distribution on the plane orthogonal to the camera's line of sight (roughly parallel to the high magnetic fiel gradient axis of the MOT). We investigated the impact of this projection with numerically-generated 3D images, and found that it had only a weak impact on both structure functions and scaling exponents. Unsurprisingly, the projected 2D images are smoother than the initial 3D images, yielding slightly larger values of $\xi_p$ in 2D than in 3D. Due to all these effects, the scaling ranges that we are able to observe are rather limited, usually below one decade, and the extracted scaling exponents can not safely be expected to be universal. However, in the present work we are mostly concerned with the relative change of the measured exponents with experimental parameters, with the aim of separating different instability regimes.

Fig.2 shows examples of SF (note the double logarithmic scale) with $p = 7.1$, for the data of Fig.1. For better comparison, the horizontal scale is normalized to $R$ for each data set and the SF curves have been shifted vertically. In most cases the SF are concave, with the highest slope for low $\ell$. When $\ell$ approaches the cloud's diameter $2R$, the SF reaches a maximum and then decreases, limiting the scaling analysis to $\ell < 2R$. We see that at low $\nabla B$ (stars), one can clearly identify two distinct scaling ranges delimited by the vertical dashed lines, one for low $\ell$ values and the other for high $\ell$. This feature is absent at higher $\nabla B$ (dots), where only a low-$\ell$ scaling range is observed. The SF corresponding to the stable MOT (circles) is very different: it is convex at low $\ell$, with a large scaling range at higher $\ell$. These behaviors will be discussed in the following.

\section{Results and Discussion}

Interesting information can be obtained from the curves $\xi_p(p)$. We present such curves in Fig.3, corresponding to the data displayed in Fig.1. The different curves correspond to different fitting ranges of $\ell$: low $\ell$ ($2 \leq \ell \leq 10$, circles), intermediate-low $\ell$ ($0.1R \leq \ell \leq 0.3R$, dots), intermediate-high $\ell$ ($0.3R \leq \ell \leq 0.6R$, squares), and high $\ell$ ($0.6R \leq \ell \leq 1.2R$, stars). The dotted line of slope 1 corresponds to the limit of a smooth field $\xi_p = p$.

We see that for unstable clouds (panels A and B), the observed behaviors are qualitatively relatively similar. The curves $\xi_p(p)$ are concave (except the low-$\ell$ ones), which is typical of intermittent turbulence. The low-$\ell$ exponent (circles) grows almost linearly with $p$ with rather large slopes (2/3 for A, and 0.78 for B), which are clearly incompatible with the KOC scaling (1/3). On the contrary, these slopes are rather close to the smooth field limit, which could indicate the presence of dissipation at small scales smearing out spatial structures. As discussed before, smoothing due to atomic motion might also play a role, especially in case B. An important difference between cases A and B is that in A the last two curves (squares and stars) are very similar, which correspond to the high-$\ell$ scaling range observed in Fig.2 (stars). The corresponding slopes are very small, consistent with the presence of shocks ($\xi_p = $const). In Fig3B, all curves are distinct which reflects the absence of a clear scaling range (except at low $\ell$).

The curves corresponding to the stable cloud, shown in panel C, are strikingly different. Here, the low-$\ell$ $\xi_p(p)$ curve has a small slope, consistent with the presence of shocks. All the other curves are relatively linear with large slopes, and lying close to each other. This indicates a large scaling range for intermediate and large $\ell$s (clearly seen in Fig.2 for the curve in circles), with a behavior close to that of a smooth field. Indeed, the fluorescence images of the stable cloud are rather smooth (see Fig.1C). However, a closer inspection reveals the presence of small-wavelength fringes, likely due to interferences between MOT beams. These small-scale ripples are responsible for the shock-like behavior of the SF at low $\ell$.

We now concentrate on the high $\ell$ scaling range that can be observed in the curve with stars in Fig.2, corresponding roughly to $0.4R \leq \ell \leq 3R$. Despite the fact that no universal behavior can be expected in this range, we can clearly see a qualitative change in the SF shape there as $\nabla B$ is varied (compare with the curve in dots). 
To quantify the presence of the high $\ell$ scaling range, we compute the difference $\Delta \xi_p$ between exponents obtained for $0.5R \leq \ell \leq R$ and $R \leq \ell \leq 2R$, normalized to the mean value. If this quantity is small then the two exponents are similar and one can speak of a scaling range for $0.5R \leq \ell \leq 2R$. In Fig.4, we plot $\Delta \xi_p$ versus $\nabla B$, which is varied between 1.2 and 12 G/cm (the MOT detuning is kept fixed at -$\Gamma$). We observe in Fig.4 two different behaviors: for $\nabla B \leq 2.4$ G/cm (upper panel), $\Delta \xi_p$ is relatively small and decreases when $p$ increases, reflecting the widening of the scaling range. The case $\nabla B = 2.4$ G/cm constitutes the limit of this regime, with $\Delta \xi_p$ quite large but still decreasing with increasing $p$ (for $p$ large). Visually (see insets), this regime corresponds to our most turbulent-looking clouds where we observe relatively small-scale structures with a complex dynamics (see Fig1A). For $\nabla B > 2.4$ G/cm (lower panel in Fig.4), the behavior is the opposite: $\Delta \xi_p$ is large and increases with $p$. There is no significant scaling range in the considered range of $\ell$. In this second regime, the cloud undergoes large deformations with spatial scales of the order of the cloud size.

\begin{figure}
\begin{center}
\includegraphics[width=1.0\columnwidth]{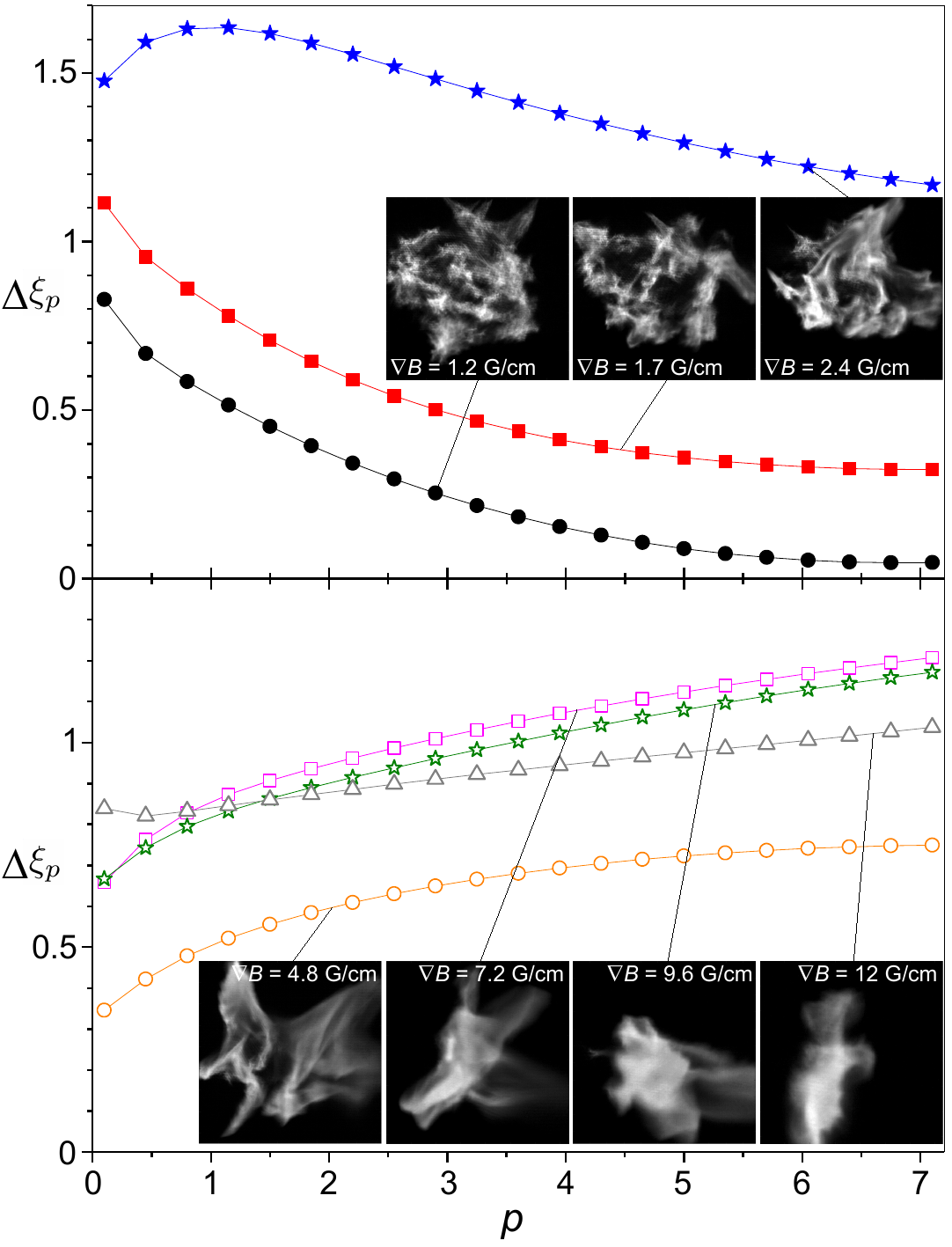}
\caption{Separation of MOT instability regimes based on structure function analysis. We plot for all unstable data sets the slope difference $\Delta \xi_p$ (see text) versus \textit{p}. Upper panel: $\nabla B = 1.2$ G/cm (black dots), $\nabla B = 1.7$ G/cm (red squares), $\nabla B = 2.4$ G/cm (blue stars); lower panel: $\nabla B = 4.8$ G/cm (orange circles), $\nabla B = 7.2$ G/cm (magenta open squares), $\nabla B = 9.6$ G/cm (green open stars), and $\nabla B = 12$ G/cm (grey open triangles). The detuning is fixed at $\delta = -\Gamma$. The ''turbulent'' regime, observed for small $\nabla B$ values, is characterized by a decrease of $\Delta \xi_p$ as $p$ increases (upper panel), while it is the opposite for the high-$\nabla B$ regime (lower panel).}
\label{fig4}
\end{center}
\end{figure}

We see that this analysis allows us to discriminate between two unstable MOT regimes, corresponding to different ranges for the magnetic field gradient. The stable regime is also easily identified. Thus, the SF-based characterization confirms the result of the visual inspection of MOT behaviors. It also agrees with the more quantitative analysis described in Ref.~\cite{Gaudesius2021}, where we identified markedly different regimes for low and high magnetic field gradients, respectively termed the ''turbulent'' and ''statistically isotropic'' regimes. A third regime (the ''anisotropic'' regime) was also observed in Ref.~\cite{Gaudesius2021}, but doesn't clearly show up in the present analysis.

Although it is difficult to directly compare experimental results from different setups, we note that our data set B (Figs.1,2 and 3) is obtained for MOT parameters approximately corresponding to those used in Ref.~\cite{Giampaoli2021}, however with different atom number. It would be interesting to compare our measured scaling exponents to those expected from the photon bubble model~\cite{Mendonca2012}.

\section{Conclusion}
In this paper we have applied the methods borrowed from turbulence to the analysis of various stable and unstable regimes of large clouds of cold atoms contained in a MOT. We have found that the structure function of the fluorescence intensity field provides an efficient tool for the classification of different dynamical regimes.

In the case of the most ''visually turbulent'' regime (low $\nabla$B), we observed two scaling ranges. At small $\ell$,  the scaling exponents $\xi_p (p)$ behave consistently with a smooth field, indicating dissipation. At intermediate $\ell$, the observed behavior of $\xi_p (p)$ is more or less consistent with KOC-type scaling for small $p$ and with the presence of shocks for larger $p$, similar to what was reported in Ref.~\cite{Iyer2018} for 3D scalar turbulence.

One of the experimental limitation in the present work is the relatively small extent of the observed scaling ranges. In the future, this could be improved by using several magnifications to image the cloud. The impact of the 3D to 2D projection and that of multiple light scattering could be reduced by using a sheet of light for imaging~\cite{Giampaoli2021} and a detuned laser~\cite{Camara2014}, provided that noise levels remain low enough. This should allow better measurements in the low-$\ell$ range.

To convincingly claim the observation of turbulence, however, one ideally needs a comparison with a simple but reasonably accurate model of MOT physics. An analytical model seems out of reach, but we have developed a 3D numerical model~\cite{Gaudesius2022} based on well-known atom-light interaction ingredients which has proven reliable for the prediction of MOT instability thresholds~\cite{Gaudesius2020} and dynamics~\cite{Gaudesius2021}. This model could be used to check the presence of turbulence and allow a comparison with the experimental data.  

\section{Acknowledgments}
This work was performed in the framework of the European Training Network  ColOpt, which is funded by the European Union (EU) Horizon 2020 program under the Marie Sklodowska-Curie action, grant No. 721465, and in the framework of the European project ANDLICA, ERC Advanced grant No. 832219. 
Adam Griffin and Sergey Nazarenko were supported by the Chaire d'Excellence awarded by
Universit\'e de la C\^ote d'Azur, France, the EU Horizon 2020 program MSC HALT project, grant No. 823937, the FET Flagships PhoQuS project, grant No. 820392, and the Simons Foundation project Collaboration on Wave Turbulence, award ID 651471.


\begin{thebibliography}{99}

\bibitem{Dyachenko1992} S. Dyachenko, A. C. Newell, A. Pushkarev, V. E. Zakharov, \textit{Optical turbulence: weak turbulence, condensates and collapsing filaments in the nonlinear Schrödinger equation}, Physica D: Nonlinear Phenomena \textbf{57}, 96 (1992).

\bibitem{Bagnato2009} E. A. L. Henn, J. A. Seman, G. Roati, K. M. F. Magalhães, and V. S. Bagnato, \textit{Emergence of Turbulence in an Oscillating Bose-Einstein Condensate}, Phys. Rev. Lett. \textbf{103}, 045301 (2009).

\bibitem{Anderson2013} T. W. Neely, A. S. Bradley, E. C. Samson, S. J. Rooney, E. M. Wright, K. J. H. Law, R. Carretero-González, P. G. Kevrekidis, M. J. Davis, and B. P. Anderson \textit{Characteristics of Two-Dimensional Quantum Turbulence in a Compressible Superfluid}, Phys. Rev. Lett. \textbf{111}, 235301 (2013).

\bibitem{Wilkowski2000} D. Wilkowski, J. Ringot, D. Hennequin, and J.-C. Garreau, \textit{Instabilities in a Magneto-optical Trap: Noise-Induced Dynamics in an Atomic System}, Phys. Rev. Lett. \textbf{85}, 1839 (2000).

\bibitem{diStefano2003} A. di Stefano, M. Fauquembergue, P. Verkerk, and D. Hennequin, \textit{Giant oscillations in a magneto-optical trap}, Phys. Rev. A \textbf{67}, 033404 (2003).

\bibitem{Labeyrie2006} G. Labeyrie, F. Michaud, and R. Kaiser, \textit{Self-sustained oscillation in a large cloud of cold atoms}, Phys. Rev. Lett. \textbf{96}, 023003 (2006).

\bibitem{Romain2016} R. Romain, A. Jallageas, P. Verkerk, and D. Hennequin \textit{Spatial instabilities in a cloud of cold atoms}, Phys. Rev. E \textbf{94}, 052212 (2016).

\bibitem{Gaudesius2020} M. Gaudesius, R. Kaiser, G. Labeyrie, Y.-C. Zhang, and T. Pohl, \textit{Instability threshold in a large balanced magneto-optical trap}, Phys. Rev. A \textbf{101}, 053626 (2020).

\bibitem{Gaudesius2021} M. Gaudesius, R. Kaiser, G. Labeyrie, Y.-C. Zhang, and T. Pohl, \textit{Phase diagram of spatiotemporal instabilities in a large magneto-optical trap}, Phys. Rev. A \textbf{103}, L053626 (2021).

\bibitem{Giampaoli2021} R. Giampaoli, J. D. Rodrigues, J. A. Rodrigues, J. T. Mendonça, \textit{Photon bubble turbulence in cold atom gases}, Nat. Commun. \textbf{12}, 3240 (2021).

\bibitem{Pohl2006} T. Pohl, G. Labeyrie, and R. Kaiser, \textit{Self-driven nonlinear dynamics in magneto-optical traps}, Phys. Rev. A \textbf{74}, 023409 (2006).

\bibitem{Stefano2004} A. di Stefano, Ph. Verkerk, and D. Hennequin, \textit{Deterministic instabilities in the magneto-optical trap}, Eur. Phys. J. D \textbf{30}, 243 (2004).

\bibitem{Mendonca2008} J. T. Mendonca, R. Kaiser, H. Tercas, J. Loureiro, \textit{Collective oscillations in ultra-cold atomic gas}, Phys. Rev. A \textbf{78}, 013408 (2008).

\bibitem{Mendonca2012} J. T. Mendonca and R. Kaiser, \textit{Photon bubbles in ultracold matter}, Phys. Rev. Lett. \textbf{108}, 033001 (2012).

\bibitem{Frisch1995} Uriel Frisch, \textit{Turbulence: The Legacy of A. N. Kolmogorov}, Cambridge University Press (1995).

\bibitem{Kolmogorov1941} A. N. Kolmogorov, \textit{The Local Structure of Turbulence in Incompressible Viscous Fluid for Very Large Reynolds Numbers}, Doklady Akademii Nauk SSSR \textbf{30}, 301 (1941).

\bibitem{Obukhov1949} A. M. Obukhov, \textit{Structure of the temperature field in turbulent flows}, Izv. Akad. Nauk SSSR, Ser. Geogr. Geofiz. \textbf{13}, 58 (1949). 

\bibitem{Corrsin1951} S. Corrsin, \textit{On the spectrum of isotropic temperature fluctuations in an isotropicturbulence}, J. Appl. Phys. \textbf{22}, 469 (1951).

\bibitem{Leveque1999} E. Leveque, G. Ruiz-Chavarria, C. Baudet, and S. Ciliberto, \textit{Scaling laws for the turbulent mixing of a passive scalar in the wake of a cylinder}, Phys. Fluids \textbf{11}, 1869 (1999).

\bibitem{Iyer2018} K. P. Iyer, J. Schumacher, K. R. Sreenivasan and P. K. Yeung, \textit{Steep Cliffs and Saturated Exponents in Three-Dimensional Scalar Turbulence},
Phys. Rev. Lett. \textbf{121}, 264501 (2018).

\bibitem{Camara2014} A. Camara, R. Kaiser, and G. Labeyrie, \textit{Behavior of a very large magneto-optical trap}, Phys. Rev. A \textbf{90}, 063404 (2014).

\bibitem{Rodrigues2022} J. D. Rodrigues, R. Giampaoli, J. A. Rodrigues, A. V. Ferreira, H. Terças, J. T. Mendonça, \textit{Quasi-static and dynamic photon bubbles in cold atom clouds}, Atoms \textit{10}, 45 (2022).

\bibitem{Gaudesius2022} M. Gaudesius, Y.-C. Zhang, T. Pohl, R. Kaiser, and G. Labeyrie, \textit{Three-dimensional simulations of spatiotemporal instabilities in a magneto-optical trap}, Phys. Rev. A \textbf{105}, 013112 (2022).

\end{thebibliography}
\end{document}